\documentclass[useAMS,usenatbib]{mn2e}

\usepackage{amsmath}
\usepackage{comment}
\usepackage{graphicx}
\usepackage{rotating}
\usepackage{color}
\usepackage{aas_macros}
\newcommand{\beq}{\begin{equation}}
\newcommand{\enq}{\end{equation}}
\newcommand{\beqa}{\begin{eqnarray}}
\newcommand{\enqa}{\end{eqnarray}}
\newcommand{\beit}{\begin{itemize}}
\newcommand{\enit}{\end{itemize}}
\newcommand{\bem}{\begin{pmatrix}}
\newcommand{\enm}{\end{pmatrix}}

\newcommand{\veck}{\mathbf{k} }

\newcommand{\lat}{\left\langle}
\newcommand{\rat}{\right\rangle}
\newcommand{\av}[1]{\lat #1 \rat}

\newcommand{\lb}{\left [}
\newcommand{\rb}{\right ]}
\newcommand{\lp}{\left (}
\newcommand{\rp}{\right )}

\newcommand{\Fab}{F_{\alpha\beta}}
\newcommand{\bes}{\begin{sideways}}
\newcommand{\ees}{\end{sideways}}

\title[Non-linear transformations for large scale structure]{Optimal non-linear transformations for large scale structure statistics}
\author[Carron and Szapudi]{J. Carron\thanks{E-mail:
carron@ifa.hawaii.edu} and I. Szapudi  \\
Institute for Astronomy, University of Hawaii, 2680 Woodlawn Drive, Honolulu, HI, 96822}
\begin{document}

\date{\today}

\pagerange{\pageref{firstpage}--\pageref{lastpage}} \pubyear{2013}

\maketitle

\label{firstpage}

\begin{abstract}

Recently, several studies proposed non-linear transformations,
such as a logarithmic or Gaussianization transformation, as 
efficient tools to recapture information about the (Gaussian) initial 
conditions. During non-linear evolution, part of the cosmologically
relevant  information leaks out from the second moment of the distribution.
This information is accessible only through complex higher order moments or, in the worst
case, becomes inaccessible to the hierarchy. The focus of this work is to investigate
these transformations in the framework of Fisher information using cosmological perturbation theory of the matter field
with Gaussian initial conditions. We show that at each order in perturbation theory, there is a
polynomial of corresponding order exhausting the information
on a given parameter.
This polynomial can be interpreted as the Taylor expansion
of the maximally efficient ``sufficient'' observable in the non-linear regime. 
We determine explicitly this maximally efficient observable for local
transformations. Remarkably, this optimal transform is essentially
the simple power transform with an exponent related to
the slope of the power spectrum; when this is $-1$, it is 
indistinguishable from the logarithmic transform. This transform Gaussianizes the distribution, and recovers the linear density contrast. Thus a direct connection is revealed between undoing of the non-linear
dynamics and the efficient capture of Fisher information. Our analytical results were compared with measurements from
the Millennium Simulation density field. We found that our transforms remain very close to optimal even in the 
deeply non-linear regime with $\sigma^2 \sim 10$.
\end{abstract}

\section{Introduction}
The non-linear regime of structure formation in the Universe is rich in cosmological information, although the extraction of this information is a serious challenge. Traditional observables in galaxy or weak-lensing surveys, such as power spectra or two-point correlation functions, are optimal in the linear, Gaussian regime, but their statistical power decreases due to the emergence of correlations between Fourier modes \citep{Meiksin:1999fk,2005MNRAS.360L..82R,2006MNRAS.370L..66N} from non-linear dynamics. The long non-linear tails in the distribution
of density fluctuations and the corresponding cosmic variance reduce the ability of observables based on moments of the field to capture the information efficiently. It has been suggested and tested with numerical simulations that non-linear transformations of the field, such as a logarithmic or a Gaussianizing map, are able to capture more efficiently this information \citep{2009ApJ...698L..90N,2011ApJ...742...91N,2011ApJ...729L..11S,2011PhRvD..84b3523Y,Joachimi:2011dq,Seo:2012bh,Carron:2012fk}, at least in the high signal to noise regime.
This effect is magnified to dramatic extent in the lognormal model of the density field \citep{1991MNRAS.248....1C}. In this case it can be shown that a large fraction of the information escapes entirely the hierarchy of  $N$-point moments in the large variance regime \citep{2011ApJ...738...86C,Carron:2012qf}. 
\newline
\newline
Our principal aim is to build an analytic theory to quantify the ability and optimality of these transforms to capture information within the matter fluctuation field. To move beyond phenomenological models or simulations, we will use cosmological perturbation theory \citep{2002PhR...367....1B}. 
Information is of course a broad concept, and optimality must refer to some simple criteria. When it comes to inference on model parameters, the ideal measure is the Fisher information. The field possesses some definite Fisher information and in the linear, Gaussian regime, this total information content is given by the ubiquitous Fisher matrix for Gaussian variables \citep{1996ApJ...465...34V,1997ApJ...480...22T}. This information is itself entirely contained within the two-point statistics of the field. Our goal is to investigate how this simple situation changes in the weakly non-linear regime.
\newline
\newline
As discussed below in some detail,  Fisher information efficient observables generically strongly depend on the model parameters of interest. Presently, we restrict our investigations to local transformations.  In that case, it is enough to perform the analysis of the information content of the one-point probability density function $p(\delta)$ and to determine how to capture this information efficiently with generic observables $\av{f(\delta)}$. This restriction simplifies drastically the analysis and singles out the variance of the fluctuations $\sigma^2$ as the sole parameter of relevance. Nevertheless, several qualitative conclusions of this work on the information in the quasi-linear regime depends only on the structure of the correlations induced by gravity with Gaussian initial conditions,
\beq \label{xiN}
\xi_N \propto \xi_2^{N-1} + \textrm{loop corrections},
\enq
and will remain unchanged in the case of a spatially correlated 
random field. Quantitative results in the correlated case are more involved and left for future work. In the case of the one-point probability distribution, the cumulants $\av{\delta}_c$ are given by
\beq \label{delc}
\av{\delta^n}_c = S_n\sigma_L^{2(n-1)} + \textrm{ loop corrections},
\enq
where $\sigma_L^2$ is the linear variance and the loop corrections involve only even powers of $\sigma_L$.
In the hierarchical model, $\sigma_L$ is identified with the variance $\sigma$ and the parameters $S_n$ are constants.  
Throughout the text, the explicit expansion parameter is the non-linear, true variance of $\delta$, which makes the notation much simpler, without changing our conclusions.
\newline
\newline
In section \ref{Efficient observables}, after introducing notations and definitions we discuss in general terms the information efficient observables, and their explicit form in terms of $p(\delta)$. In section \ref{struct}, we show that the form \eqref{xiN} of the moments induce a simple structure within the Fisher information content of the field. We discuss how to make use of this structure to obtain the observables exhausting this information, and present the leading component of this observable. In section \ref{discussion} we test our findings on the Millennium Simulation density field $p(\delta)$. We end with a discussion in section \ref{finalsection}. A set of appendices collect technical details that would break the 
continuity of the main text.
\section{On information efficient observables} \label{Efficient observables}
The information matrix of a set of observables $(f_1(\delta),f_2(\delta),\cdots)$ is the following
\beq \label{Cr}
\sum_{ij} \frac{\partial \av{f_i}}{\partial \alpha}\lb \Sigma ^{-1}\rb_{ij} \frac{\partial \av{f_j}}{\partial \beta} \le F_{\alpha \beta} = \av{ \frac{\partial \ln p}{\partial \alpha}  \frac{\partial \ln p}{\partial \beta} }
\enq
where $\Sigma_{ij} = \av{f_if_j} - \av{f_i} \av{f_j}$ is the covariance matrix, $F_{\alpha\beta}$ is the total Fisher information content of $p(\delta)$ on the parameters of interests, and the inequality is the Cram\'er-Rao inequality (an inequality between positive definite quadratic forms). In the perturbative regime, it is reasonable to expect that the information of $p$ is the same as that of the moment series. We have in that case
\beq \label{completeness}
\begin{split}
F_{\alpha \beta} &= \lim_{N \rightarrow \infty} \sum_{i,j = 1}^N\frac{\partial m_i}{\partial \alpha} \lb \Sigma^{-1}_N\rb_{ij} \frac{\partial m_j} {\partial \beta}, \quad m_i = \av{\delta^i}
\end{split}
\enq
an equality that we assume throughout the analytical part of this work. This is equivalent to assume that the functions $\partial_\alpha \ln p(\delta)$ can be expanded in powers of $\delta$ over the full range of $p$ \citep{2011ApJ...738...86C,Carron:2012qf}. While this assumption appears justified, its validity lies in the (unobservable) decay rate of $p(\delta)$ at infinity: it must not be much shallower than exponential. Careful study of the analyticity properties of the moment generating function of $p(\delta)$ in \cite{2002A&A...382..412V} and their impact on the tail suggests that strict equality does not hold in \eqref{completeness}, as for the lognormal distribution, if the slope $n$ of the power spectrum $P(k) \propto k^n$ is negative enough. As long as $\sigma$ is small enough, the mismatch in \eqref{completeness} should be negligible for all practical purposes.
\newline
\newline
Our principal results are based on the assertion that for any model parameter $\alpha$ it is always possible, at least in principle, to design a single observable that will exhaust the Fisher information of the data tied to $\alpha$, represented by the row $\Fab$. Likewise, it is always formally possible to design a single observable that contains as much information as the entire moment hierarchy.
\newline
\newline
Note that the above statement is significantly stronger than the more familiar assertion that in the family of Gaussian fields the power spectrum or two-point function exhausts the information on any parameters. The spectrum is a large collection of observables, one for each Fourier mode. On the contrary, we state that inference on a given number of parameters can be performed optimally with only the same number of observables. These observables are fine-tuned to the particular set of parameters, unlike the power spectrum.
\subsection{Formal considerations}
We can give the explicit form of the advertised 
designer or sufficient observables as follows. 
Consider the observable $o_\alpha(\delta) = \partial_\alpha \ln p(\delta)$, with the understanding that $o_\alpha$ is a function of $\delta$ only, evaluated at the fiducial value of the parameter space. By definition, we have that
\beq \label{fm}
\frac{\partial \av{o_\alpha(\delta)}}{\partial \beta} 
= \Fab,
\enq
This equality holds because $o_\alpha$ is treated as a constant in parameter space and $\partial_\alpha p = p\:\partial_\alpha \ln p$. Similarly, $o_\alpha(\delta)$ has zero mean, and its variance is then
\beq \label{var}
\av{o^2_\alpha(\delta)} - \av{o_\alpha(\delta)}^2 =  F_{\alpha \alpha}
\enq
by definition of the Fisher information. All in all, the information content of $o_\alpha$ is thus
\beq
\frac{1}{\av{o^2_\alpha(\delta)} - \av{o_\alpha(\delta)}^2}\frac{\partial \av{o_\alpha(\delta)} }{\partial \alpha}\frac{\partial \av{o_\alpha(\delta)} }{\partial \beta} = \Fab.
\enq
We recovered the corresponding row of the Fisher matrix, showing that $o_\alpha(\delta)$ is a maximally efficient observable for $\alpha$. Note that we can rescale $o_\alpha$ by a constant factor or add a constant term at our discretion. These operations do not affect the information content.
\newline
\newline
For instance, for a zero-mean single Gaussian variable $\delta$, an optimal observable is of course
\beq \label{optG}
o_\alpha(\delta) = \delta^2.
\enq
In the case of a lognormal variable, for which $\ln p \propto (\ln (1 + \delta) - \ln (1 + \sigma^2)/2)^2 /\ln(1 + \sigma^2) + $ irrelevant terms, one can set
\beq \label{optlnG}
o_\alpha(\delta) = \ln^2(1 + \delta).
\enq
These two examples are in fact rather special for two reasons. First, both optimal observables \eqref{optG} and \eqref{optlnG} could be chosen to be completely independent of the fiducial model. No fine-tuning of the observable is required at all. Second, they are both the square of a Gaussian variable. These properties are however not generic, in one or any number of dimensions. For example, for the correlated zero mean Gaussian field with spectrum $P(k)$ one has
\beq
o_\alpha(\delta) =  \int \frac{d^3k}{(2\pi)^3} \frac{|\delta(k)|^2}{P(k)} \frac{\partial \ln P(k)}{\partial \alpha},
\enq
an expression that requires prior knowledge of both the spectrum and its derivative, combining these two ingredients in an optimal filter.
Since we assume that the information is within the moment hierarchy, we can write $o_\alpha$ as a power series. Its exact representation is then
\beq \label{Mom}
o_\alpha(\delta) = \lim_{N \rightarrow \infty}\sum_{n,m = 1}^N\delta^n \lb\Sigma_N^{-1} \rb_{nm} \frac{\partial m_m}{\partial \alpha}.
\enq
Indeed, a short calculation can recover properties \eqref{fm} and \eqref{var} from this expression, with the understanding that $F_{\alpha\beta}$ is now the right hand side of \eqref{completeness}.
A more useful representation is the expansion of \eqref{Mom} in the orthogonal system associated with the moments.
It holds
\beq \label{poly}
o_\alpha(\delta)= \sum_{n = 1}^\infty s_n^\alpha P_n(\delta),
\enq
where $P_n$ is the orthogonal polynomial of degree $n$ associated
with $p(\delta)$, i.e. $\av{P_nP_m} = \delta_{nm}$, and
\beq \label{poly2}
s_n ^\alpha = \av{\frac{\partial  \ln p}{\partial \alpha} P_n}.
\enq
The total information is then
\beq \label{poly3}
F_{\alpha \beta} = \sum_{n = 1}^\infty s_n^\alpha s_n^\beta
\enq
where $s_n^\alpha s_n^\beta$ is the independent contribution of the $n$th moment to the information.
\subsection{The linear density contrast as a sufficient statistic}
Let us make an educated guess on the plausible form of our optimal observable. It is well known that the saddle-point approximation to $p(\delta)$ is of the form \citep{1994A&A...291..697B,2002PhR...367....1B}
\beq
\ln p(\delta) =  -\frac{\tau^2(\delta)}{2\sigma^2} -\frac 12 \ln \sigma^2 + \textrm{c}(\delta),
\enq
where $-\tau$ is the linear density contrast,
and $c(\delta)$ collects terms that do not depend on $\sigma^2$ and are thus irrelevant for the information.
 It follows that
\beq
\frac{\partial \ln p}{\partial \alpha} = \frac 12 \frac{\partial \ln \sigma^2}{\partial \alpha} \lp \frac{\tau^2(\delta)}{\sigma^2} - 1 \rp
\enq
up to irrelevant constants. Thus, under this approximation, an optimal observable is simply given by
\beq
o_\alpha(\delta) = \tau^2(\delta)
\enq
where we used our freedom to subtract constants and rescale by constants. The optimal transform is then simply the mapping recovering the linear density contrast.  We will be interested in power-law spectra $P(k) \propto k^n$. In this case, using the approximate form $\mathcal G(\delta) = \lp 1 + \frac{2\tau}{3}\rp^{-3/2} -1$ for the vertex generating function and the relation $\mathcal G_\delta^{\mathcal S}(\tau) = \mathcal G_\delta \lp \tau \lb 1 + \mathcal G_\delta^{\mathcal S}\rb^{-(n+3)/6} \rp$ \citep{1994A&A...291..697B} implementing smoothing effects, one has
\beq \label{linear}
 \tau(\delta) = \frac 32 \lp 1 + \delta\rp^{(n+3)/6} \lb \lp 1 + \delta \rp^{-2/3} -1 \rb.
\enq  Just as our simple examples above, the optimal observable can be chosen as the square of a Gaussian variable, and independently of the variance of the field. This illustrates in a very explicit manner the tight connection between Gaussianization, Fisher information and undoing the non-linear dynamics. In fact, this simple argument will turn out to be remarkably successful.
\section{Information in the quasi-linear regime} \label{struct}
A brute force approach presents itself to 
explicitly determine the observable. We can expand directly expressions \eqref{completeness}, \eqref{Mom} or \eqref{poly} in powers of the variance. This is possible, though rather tedious and not especially enlightening. Since the optimal observables can in principle be read out from $\partial_\alpha \ln p$, the most convenient approach, exposed in the following, turns out to be the Edgeworth series of the logarithm $\ln p$ of the probability density function. This might be surprising at first. It is well known that the Edgeworth series does not necessarily converge, and if truncated at a finite order it might not even be a sensible probability density function. However, the perturbation series obtained in that way for the quantities of interest are exact, identical to those obtained from the moments, equations  \eqref{completeness}, \eqref{Mom} and \eqref{poly}. Indeed, regardless of the question of its convergence or of the behavior of the probability density it represents, the Edgeworth series produces the correct series of moments by construction. Since we are assuming the information to be entirely within the moments, this is the only relevant property of the series. It should be viewed as a formal generating function in the case of divergence. The only relevant consequence of a divergence of the Edgeworth series is that the true total information $F$, that makes reference to the exact $p(\delta)$, might be higher than that given by that of the moment series. This is discussed in more detail in Appendix \ref{pert. series}. 
\newline
\newline
From now on we are concerned with the parameter $\ln \sigma^2$ only. The Fisher information on derived parameters is given by
\beq
F_{\alpha \beta} = \frac{\partial \ln \sigma^2}{\partial \alpha}\frac{\partial \ln \sigma^2}{\partial \beta} F
\enq
with $F = F_{\alpha= \ln \sigma^2, \beta = \ln \sigma^2}$. Due to the Gaussian initial conditions, we have
\beq
F = \frac 12 +  \textrm{corrections}.
\enq

\newcommand{\vecm}{\mathbf m}
\subsection{Structure of the information} \label{sections}
It is remarkable, and of key significance for our purpose, that the Edgeworth series of $\ln p(\delta)$ is in fact much simpler than that of $p(\delta)$. Recall that the general form of the expansion reads \citep[see also appendix \ref{pert. series}]{1998A&AS..130..193B}
\beq \label{Edgeworth}
p(\delta) =  \frac{1}{\sqrt{2\pi \sigma^2}} e^{  - \nu^2/2 } \lp1 + \Delta p(\sigma,\nu) \rp
\enq
with $\nu = \delta/\sigma$, where formally
\beq
\begin{split}
\Delta p(\sigma,\nu) &= \sum_{n = 1}^\infty \sigma^n \sum_{\veck} H_{n + 2|\veck|}(\nu) \prod_i \lp \frac{S^{k_i}_i }{i!^{k_i}k_i!} \rp \\
&=   \sigma\frac {S_3} 6 H_3(\nu) + \sigma^2 \lp \frac {S_4}{24} H_4(\nu) + \frac{S_3^2}{72} H_6(\nu)\rp + \cdots
\end{split}
\enq
The second sum runs over all set of positive integers $\veck = (k_3,k_4 \cdots)$ such that $\sum_i(i-2)k_i = k_3 + 2k_4  + \cdots= n$, and $|\veck| = \sum_i k_i$. The polynomials $H_n(\nu)$ are the Hermite polynomials. At each power $n$ of the variance, the polynomial of lowest degree is $H_{n + 2}$ and that of highest degree is $H_{3n}$. One should therefore naively expect the term of order $\sigma^n$ of $\ln p$ to be a polynomial of degree $3n$ as well. This would mean that high order moments contribute very quickly to the information. It is known \citep{Takemura}, however, that remarkable cancellations occur in the expansion of $\ln p(\delta)$ producing a polynomial of degree $n + 2$, the lowest degree in the corresponding term in the expansion of $p(\delta)$. To second order we obtain from \eqref{Edgeworth}, discarding some irrelevant constants,
\beq \label{lp}
\begin{split}
\ln p &= - \frac{\nu^2}{2} - \frac 12 \ln \sigma^2 + \sigma \frac {S_3} 6 H_3(\nu) \\
&+ \sigma^2 \lb \frac {S_4}{24} H_4(\nu) + \frac{S_3^2}{36} \lp H_6(\nu) - H_3^2(\nu) \rp \rb + \cdots.
\end{split}
\enq
The term $H_6 - H_3^2$ is a polynomial of degree $4$, so that the entire $\sigma^2$ term is a polynomial of degree $4$. The expansion of $\ln p$ was studied in detail in the univariate and multivariate case by \cite{Takemura}, to which we refer for details and a proof of these cancellations. To summarize, we have formally
\beq  \label{}
\ln p =- \frac{\nu^2}{2} - \frac 12 \ln \sigma^2 + \sum_{n = 1}^\infty \sigma^n g_{n+2}(\nu)
\enq
where $g_k(\nu)$ is a polynomial of degree $k$, which, as $H_k(\nu)$, contains only powers of $\nu$ of the same parity as $k$. 
 We are interested in the derivatives of $\ln p$. We have
\beq\label{lnp}
\begin{split}
\frac{\partial \ln p}{\partial \ln \sigma^2} &= \frac 12 H_2(\nu) + \sum_{n= 1}^\infty \sigma^n \lp \frac n 2 g_{n +2}(\nu)- \frac {\nu} {2} g^{'}_{n +2}(\nu) \rp \\
&=   \sum_{n= 0}^\infty \sigma^n r_{n+2}(\nu)
\end{split}
\enq
where $r_k(\nu)$ is again some polynomial of degree $k$ that includes only powers with the same parity as $k$.
If loop corrections are taken into account, it is easy to see that the same result holds with $\sigma_L$ in place of $\sigma$ (with of course different polynomials $r_{n+2}$).
\newline
\newline
Let us discuss some immediate consequences of \eqref{lnp}.
\newline
\newline
(i) The expansion of the Fisher information matrix contains only even powers of $\sigma$. It is thus truly an expansion in powers of the variance $\sigma^2$, unlike the Edgeworth series  which is an expansion in $\sigma$. This follows directly from the parity properties of the polynomials entering $p$ and $\partial_\alpha \ln p$.
\newline
(ii) Comparing equation \eqref{lnp} to its exact form $\sum_ns_nP_n(\nu)$, we infer that the leading term of $s_{n}$ is of order $\sigma^{n-2}$. If it were lower, then $\partial_{\ln \sigma^2} \ln p$ would contain at that lower order a term proportional to $\nu^n$, but we have seen that it does not. Besides, the expansion of $s_n$ can contain only powers of the same parity as $n$. Otherwise, the expansion of the information would also contain odd powers in the variance. We see that the independent contribution $F_n = s_n^2$ of the $n$th moment to the information is $\sigma^{2n-4}$ and higher. Thus, for any $n$,
\beq
F = \sum_{i,j = 1}^{n+2} \frac{\partial m_i}{\partial \ln\sigma^2}\lb \Sigma_{n+2}^{-1} \rb_{ij}\frac{\partial m_j}{\partial \ln \sigma^2} + O\lp \sigma^{2n +2}\rp.
\enq
In other words, the $n$ first terms (the Gaussian information being the $0$th term) of the total information are entirely within the first $n+2$ moments, a rather remarkable structure not obvious at first sight.
It also follows immediately that we can always devise a polynomial of order $n+2$ that captures that entire information, neglecting terms in $\sigma^{2n+2}$ and higher. 
\newline
(iii) The form \eqref{lnp} is not limited to the case of a single variable. This property of the polynomial expansion of the joint probability $\ln p(\delta(x_1)\delta(x_2)\cdots)$ holds whenever the leading term of the joint cumulants $\xi_N$ is proportional to some expansion parameter of the adequate power,
\beq
\xi_N(\delta(x_1)\cdots\delta(x_N)) \propto \epsilon^{N-1} + \cdots
\enq
where the expansion parameter coincide with the variance in the one-dimensional case. Even though there is no such simple and explicit expansion parameter for the $N$-point functions, this is nonetheless the natural expansion in our case since  $\xi_N \propto \xi_2^{N-1} + \textrm{loop corrections}$ in perturbation theory. We can conclude that for any $n$, the information in
a spatially correlated field is as above contained within the first $n$ point functions when neglecting terms of order $2n+2$ and higher. Again, one can define a (multivariate) polynomial of order $n+2$ in the field that captures that entire information.
\newline
\newline
Let us now proceed with the discussion of the information efficient observables. We work out the general structure of these observables, assuming all series of interests do actually converge. We define $R_{nk}$ to be the coefficient of $\nu^k$ in $r_{n}(\nu)$, and similarly $G_{nk}$ is the coefficient of $\nu^k$ in $g_n(\nu)$. We reorganize the series \eqref{lnp} using $\nu = \delta/\sigma $, multiply by $2\sigma^2$ such that we will recover $\delta^2$ for the Gaussian distribution, and ignore the irrelevant constant term. Following these steps we obtain the following expression
\beq
\begin{split}
o(\delta) &= \sum_{i = 0}^\infty \sigma^i  \lp  2 \sum_{n = 1 }^{\infty} R_{n+i ,n} \delta^n \rp.
\end{split}
\enq
The series in parenthesis defines some function $f_i$ of $\delta$, so that we can write
\beq \label{exp}
o(\delta) = f_0(\delta) + \sigma^2f_2(\delta) + \sigma^4f_4(\delta) + \cdots
\enq
Only even powers occur due to the parity properties of the $r$ polynomials. The function $f_0$ is given by the leading terms of the $r$ polynomials, $f_2$ the next to leading, and so forth. Recall that $r_{n}(\nu)$ is obtained from $g_n(\nu)$ according to equation \eqref{lnp}. Component-wise, this relation gives
\beq
R_{n + 2,k} = \frac 12 \lp n - k \rp G_{n+2,k}, \quad n \ge 1.
\enq
Using this relation, in terms of the matrices $G$ the different observables $f_i$ takes the simple form
\beq \label{structure}
f_i(\delta) = (i-2) \sum_{n = 1}^\infty G_{n +i,n}\delta^n.
\enq
In particular, $f_2(\delta)$ is zero, such that
\beq
o(\delta) = f_0(\delta) + \sigma^4f_4(\delta) + \cdots.
\enq
Note that the powers of $\sigma$ in \eqref{exp} do not reflect at which order they enter the information. We will see below that $f_4(\delta)$ only contributes to $\sigma^6$. The leading component $f_0$ is
\beq
f_0(\delta) = \delta^2 - 2\sum_{n = 3}^\infty G_{nn}\delta^n.
\enq
It does contain only the leading hierarchical cumulants, and no term linear in $\delta$.
\newline
\newline
For further reference, we have directly from \eqref{lp}
\beq \label{gs}
\begin{split}
g_3(\nu) &= \frac{S_3}{6}\lp \nu^3 - 3\nu \rp \\
g_4(\nu) &=  \frac {S_4}{24} \lp \nu^4 - 6\nu^2 + 3\rp - \frac{S_3^2}{24} \lp 3\nu^4 - 12\nu^2 + 5 \rp,
\end{split}
\enq
loop corrections entering only $g_5$ and higher. In appendix \ref{appendixF} we give $g_5$ and $g_6$ within the hierarchical model.
\subsection{Constructing the optimal observable}
Given the structure displayed above, it is simple to obtain the information content including all terms below a given power of the variance, as well as the observables exhausting entirely this information. These quantities can be read out from the polynomials $g_n(\nu)$ entering $\ln p$. To obtain the information content including up to $\sigma^{2n}$, one has to keep track of the polynomials $g_3(\nu)$ up to $g_{2+2n}(\nu)$. Interestingly, to obtain the optimal observable capturing that same information, it is enough to truncate at $g_{2+n}(\nu)$. 
For instance, with the explicit form of $g_3$ and $g_4$ in equation \eqref{gs}, we can obtain the first, $\sigma^2$ term of the total information $F$, but also the observable capturing the first two terms of $F$, proportional to $\sigma^2$ and $\sigma^4$. In this respect, it is twice as simple to obtain the optimal observable than the total information content. We prove this non-trivial but convenient fact in appendix \ref{2simple}.
\newline
\newline
These considerations lead immediately to one of the main results of this paper. Reading out $g_3(\nu)$ and $g_4(\nu)$ in \eqref{gs}, we have that 
\beq
o(\delta)  = \delta^2 - \frac{S_3}{3} \delta^3 + \frac 1 {12} \lp 3S_3^3  - S_4\rp \delta^4
\enq
captures the entire information, when neglecting terms of order $\sigma^6$ and higher in the expansion of $F$.  Remarkably, to that order only $f_0(\delta)$ contributes to the information, such that the optimal observable is still {\em independent of the variance } of the field and of loop corrections. This situation changes only when interested in capturing the $\sigma^6$ term or higher in the information, where $f_4(\delta)$ or higher are necessary. We discuss $f_4(\delta)$ in appendix \ref{appendixF}, and give there its first two coefficients, proportional to $\delta$ and $\delta^2$, in the hierarchical model.
\newline
Since $f_0(\delta)$ completely dominates the information, we focus on this observable in the following. We derive in appendix \ref{appendixF}
\beq \label{f0}
\begin{split}
f_0(\delta)& = \sum_{n} a_n \delta^n\quad \textrm{with} \\
a_n  &=\frac{2}{n!} \sum_{\veck} (-1)^{|\veck| }\lp n -2+ |\veck| \rp! \prod_{i \ge 3}  \frac{S_i^{k_i}}{  \lp i-1 !\rp^{k_i}k_i!},
\end{split}
\enq
where the second sum runs over all vectors of positive integers $\veck = (k_3,k_4,\cdots)$ of any dimension such that $\sum_{i} i k_i =  n - 2 $,
and where $|\veck |$ stands for $ \sum_i k_i$. We already derived $a_n$ for $n = 0$ to $n = 4$. Further, for $n = 5$, contributing are $\veck = (0,0,1),(1,1),(3)$, and for $n = 6$, $\veck= (0,0,0,1),(1,0,1),(0,2),(2,1),(4)$. These coefficients can also be read out from $g_5$ and $g_6$ in equation \eqref{gp}. The full list of the first six Taylor coefficients are given by
\beq
\begin{split}
a_0 &= a_1 = 0, \quad a_2 = 1 \\
a_3 &= -\frac{S_3}{3},\quad a_4 = \frac{1}{12}\lp 3S_3^3 - S_4\rp \\
a_5 &= -\frac{S_5}{60}+ \frac{S_3S_4}{6}  - \frac{S_3^3}{4} \\
a_6 &= -\frac{S_6}{360} + \frac{S_3S_5}{24} + \frac{S_4^2}{36} - \frac 7 {24} S_3^3S_4 +\frac 7 {24} S_3^4.
\end{split}
\enq
\begin{figure}
\begin{center}
\includegraphics[width = 0.5\textwidth]{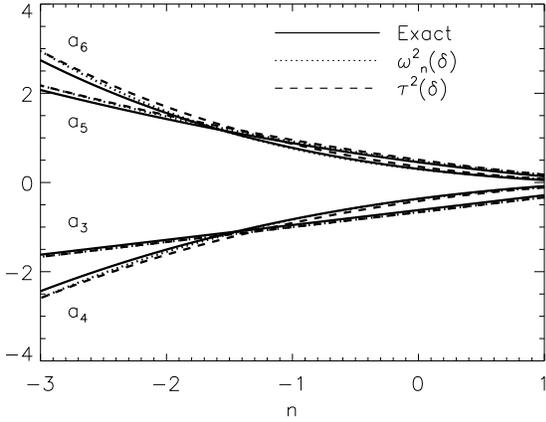}
\caption{The first coefficients of the Taylor expansion of the observable $\delta^2 + \sum_{k \ge 3}a_k\delta^k$ capturing the information in the quasi-linear regime, for power-law spectrum $P(k) \propto k^n$. The dotted lines shows the coefficients of the power transform $\omega^2_n(\delta)$ and the dashed line those of the squared linear density $\tau^2(\delta)$, as given in \eqref{pow} and \eqref{linear}. When $n = -1$, $\omega_n$ is the logarithmic mapping $\ln(1 + \delta)$.}
\label{figmap}
\end{center}
\end{figure}
\newline
\newline
 In Fig. 1 \ref{figmap} we show the value of these coefficients for a power-law spectrum $P(k) \propto k^n$, using the values of $S_3$ to $S_6$ from \cite{1994A&A...291..697B}.  It is clear from the figure that the leading optimal observable is a strong function of $n$, and the smaller $n$ is,
the stronger function of $\delta$. Remember that $f_0(\delta)$ does not contain a linear term, so that we can tentatively write it as the square of the non-linear transformation. The cumulants are very small for $n \sim 2$, and it is known that the lognormal distribution is a good match to $p(\delta)$ for $n \sim - 1$. The following observable therefore suggests itself
 \beq\label{pow}
 \omega^2_n(\delta) = \lp  \frac{\lp 1 + \delta \rp^{ (n+1)/3} -1}{(n+1)/3}\rp^2,
 \enq
 that interpolates between no transformation, $\omega_2(\delta) = \delta$ for $n = 2$, and the exact logarithmic mapping, $\omega_{-1}(\delta) = \ln(1 + \delta)$ for $n = -1$. Note that  $\omega_n(\delta)$ is simply the power (Box-Cox) transformation of $\delta$ with exponent $(n +1)/3$. The agreement between  $\omega_n^2$, shown as the dotted line on the figure and the exact coefficients is remarkable for any value $n$ of interest. As discussed earlier, one should expect according to the saddle-point approximation to $p(\delta)$ the leading observable to be $\tau^2(\delta)$, where $-\tau$ is the linear density. The dashed line on figure \ref{figmap} shows the coefficients of $\tau^2$, according to the approximation \eqref{linear}. The agreement is again excellent, confirming our expectations.
\subsection{Leading non-Gaussian information}\label{leadingF}
It is worth discussing the total information content of $p(\delta)$ in order to make a connection to previous results in the literature and to illustrate ours. One way to obtain the total information is from the decomposition \eqref{lnp} together with $F = \partial_\alpha \av{\partial_\alpha \ln p}$, with $\partial_\alpha \ln p$ fixed in parameter space. Expanding the polynomials in terms of their matrix elements gives us
\beq
F = \sum_{n,k = 0}^\infty \sigma^n R_{n + 2,k}\frac{1}{\sigma^k}\frac{\partial m_k}{\partial \ln \sigma^2}.
\enq
With $g_3$ and $g_4$ in \eqref{gs}, and $r_2 = H_2 /2$, we get
\beq
\begin{split} \label{Fs2}
F &= R_{22} + \sigma^2 \lp 6 R_{44} + 2S_3 \:R_{33}  \rp + O(\sigma^4) \\
  &= \frac 12 -\frac 14 \sigma^2 \lp S_4 - S_3^2 \rp  +  \sigma^2 \frac{S_3^2}{6} + O(\sigma^4)
\end{split}
\enq
For clarity, we separated in the last line 
the contribution to the information from the second and third moments. 
The first term, $1/2$, is the Gaussian information, the second represents the change in the covariance of the second moment due to non-Gaussianity. The presence of the kurtosis $S_4$ is expected, since it enters directly the variance of $\delta^2$. One way to understand the modulation with $-S_3^2$ is that a linear piece could be added to $\delta^2$, with zero mean but reducing its variance. 
The third term is the independent information content of the third moment, derived in the multivariate setting in \cite{2001MNRAS.328.1027T}; our second
term is a generalization to their results.
The reason for the difference is that
\cite{2001MNRAS.328.1027T} truncate the expansion of $p(\delta)$ after the first order term. We expand $F$ in powers of the variance, therefore we include all
terms up to order $\sigma^2$. As a consequence, 
in contrast to the conclusions of \cite{2001MNRAS.328.1027T}, the leading change in $F$ is not necessarily positive, but can have any sign depending on the value of the cumulants $S_3$ and $S_4$.
\newline
\newline
According to our reasoning in the last section, the observable $\delta^2 - \frac {S_3} 3 \delta^3 $ captures the entire expression in \eqref{Fs2}. This can be illustrated simply, providing us with a simple sanity check of our methods and results. Consider more generically the observable
\beq
f(\delta) = \delta^2 + a_3 \delta^3
\enq
as a function of $a_3$. We have $\av{f} = \sigma^2 + a_3 S_3 \sigma^4$. It follows
\beq
\begin{split}
 \lp \frac{\partial \av{f}}{\partial \ln \sigma^2}\rp^2 & = \sigma^4\lp 1 + 4a_3S_3\sigma^2   + O(\sigma^4)\rp \\
\av{f^2} - \av{f}^2& = m_4 + 2a_3\:m_5 + a_3^2 m_6 - \lp \sigma^2 + a_3 S_3\sigma^4\rp^2 \\
&= \sigma^4 \lb 2 + \sigma^2 \lp S_4 + 18S_3a_3 + 15a_3^2 \rp + O(\sigma^4)\rb.
\end{split}
\enq
Building the ratio, the information content of $f$ becomes
\beq
\begin{split}
&\frac 12 + \sigma^2 \lp -\frac {S_4}4 - \frac 52 a_3S_3 - \frac {15}4 a_3^2 \rp + O(\sigma^4) \\
= &\frac 12 - \frac 14 \sigma^2 \lp S_4 - S_3^2\rp + \sigma^2 \frac{S_3^2}{6} \\
&\quad - \frac {15} {4}\sigma^2 \lp a_3 + \frac {S_3}3\rp^2 + O(\sigma^4).  
\end{split}
\enq
Clearly, this expression reaches a maximum precisely when $a_3 = -S_3 /3$, when we recover the total information \eqref{Fs2}. Equation \eqref{Fs2} is still independent of loop corrections. In appendix \ref{appendixF} we give the next term in the expansion of $F$ within the hierarchical model.


\section{Tests to simulations} \label{discussion}

We used the publicly available matter density field from the
Millennium Simulation \citep{SpringelEtal2005} to estimate the information content of our optimal observables.
We calculated the probability distribution function $p(\delta)$ 
in the $z=0$ $\Lambda$CDM dark matter field of $500h^{-1} \textrm{Mpc}$ box size on a $256^3$ grid
running the cumulative grid algorithm \citep[see][for details]{Szapudi2009}
on several scales $i \times 1.95h^{-1}\textrm{Mpc}$, with $i=1\ldots 29$.
In addition, we measured moments, negative and log moments, and cumulants 
directly from the grid. Our accuracy for $p(\delta)d\delta$ was $6-8\times 10^{-8}$ for each
scale with $d\delta=0.001$, and we checked that from $p(\delta)$ we can recover the grid-direct moments,
negative and log moments to sub-percent accuracy. Note that Poisson noise was negligible even on our smallest base scale, $1.95h^{-1}\textrm{Mpc}$, with average count of about $600$ particles per cell.
\begin{figure}
\begin{center}
\includegraphics[width = 0.5\textwidth]{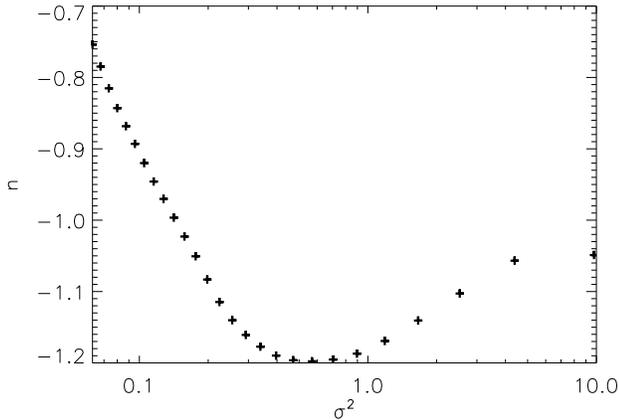}
\caption{The values of $n = -3 - \partial \ln \sigma^2 / \partial \ln R$  as measured from the Millennium Simulation density field, and used in our observables $\omega_n^2(\delta)$ and $\tau^2(\delta)$ (see Eqs \eqref{linear} and \eqref{pow}). Throughout the scales probed, $n$ never deviates substantially from $-1$, for which the optimal transformation is indistinguishable from the logarithmic mapping.}
\label{figslope}
\end{center}
\end{figure} We then obtained the derivatives $\partial_{\ln \sigma^2} \ln p(\delta)$ and the slope $n = -3-\partial_{\ln R} \ln \sigma^2$ at each scale, using finite differences. With these derivatives we evaluate the total information content $F = \av{\lp \partial_{\ln \sigma^2} \ln p\rp^2 }$ and that of of our observables, implementing straightforwardly the formulae in Eq. \eqref{Cr}.
\begin{figure}
\begin{center}
\includegraphics[width = 0.5\textwidth]{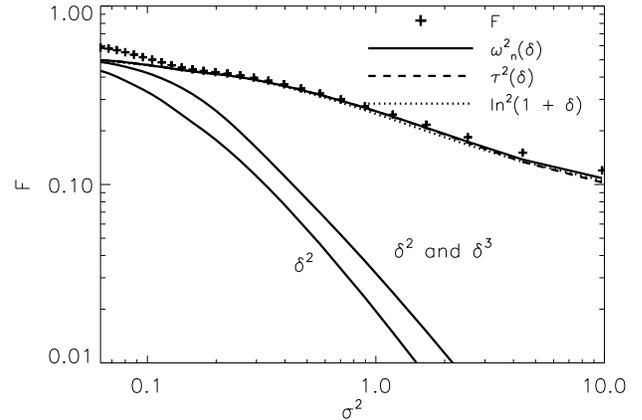}
\caption{The total information content of $p(\delta)$ on the parameter $\ln \sigma^2$ in the Millennium Simulation density field, as a function of the variance of the field (crosses), together with that of the observables $\omega_n^2(\delta), \tau^2(\delta)$ and $\ln^2(1+ \delta)$, indicated by the upper solid, dashed and dotted lines, These three observables are very close to optimal throughout the entire range probed, even in the very deeply non-linear regime. Also shown for comparison are the information content of $\delta^2$ and the combined information of $\delta^2$ and $\delta^3$ (lower solid lines). They show the steep decay characteristic of distributions with heavy tails. On the largest scales, the total information is noticeably higher than the Gaussian value $1/2$. This is due to artificial information originating from cosmic variance, entering high order statistics but not the smooth observables shown on the figure (See Section \ref{discussion} in the text).}
\label{figinfo}
\end{center}
\end{figure}
\newline
The crosses on figure \ref{figinfo} show the total information of $p(\delta)$. It might be surprising that it does not asymptote to the Gaussian value $1/2$ on our very largest scales $\sigma^2\sim 0.1$, but is in fact slightly higher, $ F\approx 0.6$. We found $p(\delta)$ had significant cosmic variance on these scales due to the relatively small volume of the simulations. 
For this reason, the shape of $p(\delta)$ contains some artificial features, deviating slightly from an exact Gaussian. These features propagate to the derivatives of $p(\delta)$. Any feature in the derivatives contributes to the total information, which is why $F$ is slightly higher than the Gaussian value. These features however do not contributes to the information within the smooth observables $\omega_n^2(\delta), \tau^2(\delta)$ and $\ln^2(1 + \delta)$, shown as the upper solid, dashed and dotted lines. These curves tend to $1/2$ as expected. It is striking, and somewhat unexpected, that these three observables remain in fact essentially optimal throughout the entire dynamic range. They still capture as much as $90\%$ of the total information when $\sigma^2 \approx 10$. On the other hand, the information of $\delta^2$ and the combination of $\delta^2$ and $\delta^3$, lower solid lines, show the sharp decay characteristic of the lognormal distribution, and are down by orders of magnitudes on our smaller scales shown $\sim 4 h^{-1}$Mpc.
Figure \ref{figslope} shows the value of $n$ measured and used at each variance. We find that $\omega_n$ and $\tau$ are essentially indistinguishable over the full range, while the logarithmic transform, independent of $n$, is only very slightly less powerful when $\sigma^2 \approx 1$, where $n$ deviates the most from $-1$.
\section{Conclusions} \label{finalsection}
Let us discuss our main results and future prospects. We presented a rigorous approach to understanding the information content of the density field evolving from Gaussian initial conditions. We described how one can obtain the maximally efficient, "sufficient", observables. We showed that the structure of the moments under the action of gravity in the quasi-linear regime makes a clear prediction of the shape of these observables, and of the associated non-linear transformations. To a very good approximation, optimal observables can be chosen independently of the variance of the field. They are for this reason fundamental observables associated to a set of hierarchical cumulants, and can be applied to data with minimal fine-tuning, a most desirable property. The optimal mapping depends on the slope $n$ of the power spectrum, coinciding with the logarithmic mapping only if $n = -1$. We found with the help of the Millenium Simulation that in practice  the slope is close enough to $-1$ for the scales of interest so that the logarithmic mapping remains essentially optimal. We established in this way a direct connection between the optimal transformation, the linear density contrast $-\tau$, the well-known logarithmic mapping $\ln (1 + \delta)$, and its generalization to other slopes, the Box-Cox transformation $\omega_n(\delta)$. The success of these transforms lies within perturbation theory. Three different facets of non-linear transforms, the undoing of the non-linear dynamics, the capture of information on parameters, and the Gaussianization of the field become unified in this picture. 
\newline
The methods we expose in this work are fairly general and we conjecture that they can be useful in a variety of situations. We have concentrated on the one-point $p(\delta)$, with the justification that we were interested in local transformations. A next logical step would be to investigate in more detail the case of a spatially correlated field, making similar use of the high level of structure displayed in $\ln p$. The required calculations are more tedious but entirely analogous to what has been presented. In fact, a first approximation for the total Fisher information can be obtained by simply multiplying our 
results with $n_p$, the number of pixels (or the number of effective pixels).
Our techniques can be relatively straightforwardly adapted to other random fields, such as weak lensing convergence, or CMB maps. The theory is general enough that optimal observables can be constructed for Poisson or sub-Poisson scatter encountered in galaxy catalogs and simulations, or to deal with practical issues such as redshift distortions, bias (especially in the context of the halo model), and projection effects for 2-dimensional surveys. Also, non-Gaussian initial conditions can be implemented in this approach. These and other possible generalizations are left for subsequent research.

\section*{Acknowledgments}
We acknowledge NASA grants NNX12AF83G and NNX10AD53G for support, 
and the Polanyi program of the Hungarian National Office for the Research and Technology (NKTH). 
We thank the reviewer Michael Vogeley for useful suggestions.


\bibliographystyle{mn2e}
\bibliography{bib}

\begin{thebibliography}{}

\bibitem[\protect\citeauthoryear{{Bernardeau}}{{Bernardeau}}{1994}]{1994A&A...%
291..697B}
{Bernardeau} F.,  1994, \aap, 291, 697

\bibitem[\protect\citeauthoryear{{Bernardeau}, {Colombi}, {Gazta{\~n}aga} \&
  {Scoccimarro}}{{Bernardeau} et~al.}{2002}]{2002PhR...367....1B}
{Bernardeau} F.,  {Colombi} S.,  {Gazta{\~n}aga} E.,    {Scoccimarro} R.,
  2002, \physrep, 367, 1

\bibitem[\protect\citeauthoryear{{Blinnikov} \& {Moessner}}{{Blinnikov} \&
  {Moessner}}{1998}]{1998A&AS..130..193B}
{Blinnikov} S.,  {Moessner} R.,  1998, \aaps, 130, 193

\bibitem[\protect\citeauthoryear{{Carron}}{{Carron}}{2011}]{2011ApJ...738...86%
C}
{Carron} J.,  2011, \apj, 738, 86

\bibitem[\protect\citeauthoryear{{Carron}}{{Carron}}{2012}]{Carron:2012fk}
{Carron} J.,  2012, Physical Review Letters, 108, 071301

\bibitem[\protect\citeauthoryear{{Carron} \& {Neyrinck}}{{Carron} \&
  {Neyrinck}}{2012}]{Carron:2012qf}
{Carron} J.,  {Neyrinck} M.~C.,  2012, \apj, 750, 28

\bibitem[\protect\citeauthoryear{{Coles} \& {Jones}}{{Coles} \&
  {Jones}}{1991}]{1991MNRAS.248....1C}
{Coles} P.,  {Jones} B.,  1991, \mnras, 248, 1

\bibitem[\protect\citeauthoryear{{Joachimi}, {Taylor} \&
  {Kiessling}}{{Joachimi} et~al.}{2011}]{Joachimi:2011dq}
{Joachimi} B.,  {Taylor} A.~N.,    {Kiessling} A.,  2011, \mnras, 418, 145

\bibitem[\protect\citeauthoryear{{Meiksin} \& {White}}{{Meiksin} \&
  {White}}{1999}]{Meiksin:1999fk}
{Meiksin} A.,  {White} M.,  1999, \mnras, 308, 1179

\bibitem[\protect\citeauthoryear{{Neyrinck}}{{Neyrinck}}{2011}]{2011ApJ...742.%
..91N}
{Neyrinck} M.~C.,  2011, \apj, 742, 91

\bibitem[\protect\citeauthoryear{{Neyrinck}, {Szapudi} \& {Rimes}}{{Neyrinck}
  et~al.}{2006}]{2006MNRAS.370L..66N}
{Neyrinck} M.~C.,  {Szapudi} I.,    {Rimes} C.~D.,  2006, \mnras, 370, L66

\bibitem[\protect\citeauthoryear{{Neyrinck}, {Szapudi} \& {Szalay}}{{Neyrinck}
  et~al.}{2009}]{2009ApJ...698L..90N}
{Neyrinck} M.~C.,  {Szapudi} I.,    {Szalay} A.~S.,  2009, \apjl, 698, L90

\bibitem[\protect\citeauthoryear{{Rimes} \& {Hamilton}}{{Rimes} \&
  {Hamilton}}{2005}]{2005MNRAS.360L..82R}
{Rimes} C.~D.,  {Hamilton} A.~J.~S.,  2005, \mnras, 360, L82

\bibitem[\protect\citeauthoryear{{Seo}, {Sato}, {Dodelson}, {Jain} \&
  {Takada}}{{Seo} et~al.}{2011}]{2011ApJ...729L..11S}
{Seo} H.-J.,  {Sato} M.,  {Dodelson} S.,  {Jain} B.,    {Takada} M.,  2011,
  \apjl, 729, L11+

\bibitem[\protect\citeauthoryear{{Seo}, {Sato}, {Takada} \& {Dodelson}}{{Seo}
  et~al.}{2012}]{Seo:2012bh}
{Seo} H.-J.,  {Sato} M.,  {Takada} M.,    {Dodelson} S.,  2012, \apj, 748, 57

\bibitem[\protect\citeauthoryear{{Springel}, {White}, {Jenkins}, {Frenk},
  {Yoshida}, {Gao}, {Navarro}, {Thacker}, {Croton}, {Helly}, {Peacock}, {Cole},
  {Thomas}, {Couchman}, {Evrard}, {Colberg} \& {Pearce}}{{Springel}
  et~al.}{2005}]{SpringelEtal2005}
{Springel} V.,  {White} S.~D.~M.,  {Jenkins} A.,  {Frenk} C.~S.,  {Yoshida} N.,
   {Gao} L.,  {Navarro} J.,  {Thacker} R.,  {Croton} D.,  {Helly} J.,
  {Peacock} J.~A.,  {Cole} S.,  {Thomas} P.,  {Couchman} H.,  {Evrard} A.,
  {Colberg} J.,    {Pearce} F.,  2005, \nat, 435, 629

\bibitem[\protect\citeauthoryear{Stewart}{Stewart}{1997}]{Stewart:1997uq}
Stewart G.~W.,  1997, IMA Journal of Numerical Analysis, 17, 1

\bibitem[\protect\citeauthoryear{{Szapudi}}{{Szapudi}}{2009}]{Szapudi2009}
{Szapudi} I.,  2009, in {Mart{\'{\i}}nez} V.~J.,  {Saar} E.,
  {Mart{\'{\i}}nez-Gonz{\'a}lez} E.,   {Pons-Border{\'{\i}}a} M.-J.,  eds, Data
  Analysis in Cosmology Vol.~665 of Lecture Notes in Physics, Berlin Springer
  Verlag, {Introduction to Higher Order Spatial Statistics in Cosmology}.
pp 457--492

\bibitem[\protect\citeauthoryear{{Takemura} \& {Takeuchi}}{{Takemura} \&
  {Takeuchi}}{1988}]{Takemura}
{Takemura} M.,  {Takeuchi} K.,  1988, Sankhya : The Indian Journal of
  Statistics, Series A, 50, 111

\bibitem[\protect\citeauthoryear{{Taylor} \& {Watts}}{{Taylor} \&
  {Watts}}{2001}]{2001MNRAS.328.1027T}
{Taylor} A.~N.,  {Watts} P.~I.~R.,  2001, \mnras, 328, 1027

\bibitem[\protect\citeauthoryear{{Tegmark}, {Taylor} \& {Heavens}}{{Tegmark}
  et~al.}{1997}]{1997ApJ...480...22T}
{Tegmark} M.,  {Taylor} A.~N.,    {Heavens} A.~F.,  1997, \apj, 480, 22

\bibitem[\protect\citeauthoryear{{Valageas}}{{Valageas}}{2002}]{2002A&A...382.%
.412V}
{Valageas} P.,  2002, \aap, 382, 412

\bibitem[\protect\citeauthoryear{{Vogeley} \& {Szalay}}{{Vogeley} \&
  {Szalay}}{1996}]{1996ApJ...465...34V}
{Vogeley} M.~S.,  {Szalay} A.~S.,  1996, \apj, 465, 34

\bibitem[\protect\citeauthoryear{{Yu}, {Zhang}, {Lin}, {Cui} \& {Fry}}{{Yu}
  et~al.}{2011}]{2011PhRvD..84b3523Y}
{Yu} Y.,  {Zhang} P.,  {Lin} W.,  {Cui} W.,    {Fry} J.~N.,  2011, \prd, 84,
  023523

\end{thebibliography}
\begin{appendix}
\section{Perturbative expansion of the optimal observable}\label{2simple}
In order to calculate the perturbation series of the Fisher information including $\sigma^{2n}$, it is necessary to keep track of the terms of that  same order in $\partial_\alpha \ln p$. Nevertheless, we used the fact in the the text that in order to read out observables capturing that same information, it is enough to truncate the series of $\partial_\alpha \ln p$ at $\sigma^n$. We prove this assertion next.
\newline
\newline
Writing schematically the expansion of $p(\delta)$ and $\partial_\alpha \ln p(\delta)$ as
\beq
p = p_G\lp 1 + \sum_{i \ge 0 }p^{(i)}\rp \quad \frac{\partial \ln p}{\partial \alpha} = \sum_{i \ge 0}\frac{\partial \ln p^{(i)}}{\partial \alpha},
\enq
consider the observable obtained by truncating the series at order $n$,
\beq
o_\alpha(\delta) = \sum_{k = 0}^n \frac{\partial \ln p^{(k)}(\delta)}{\partial \alpha}.
\enq
We have then formally
\beq
\frac{\partial \av{o}}{\partial \alpha}  = \sum_{i,j = 0}^{2n} \sum_{k = 0}^n p_{ijk},
\enq
where we discarded summation indices higher than $2n$, and $p_{ijk} = \av{p^{(i)}\lp \frac{\partial \ln p^{(j)}}{\partial \alpha}\rp\lp \frac{\partial \ln p^{(k)}}{\partial \alpha}\rp}_G$.
Similarly, neglecting such higher order terms, its variance reduces to
\beq
\av{o^{2}_\alpha(\delta)} - \av{o_\alpha(\delta)}^2 = \sum_{i = 0}^{2n}\sum_{j,k = 0}^n p_{ijk}.
\enq
Writing now
\beq
\sum_{i,j = 0}^{2n}\sum_{k = 0}^n  p_{ijk} = \sum_{i = 0}^{2n}\sum_{j,k = 0}^n  p_{ijk} \: +  \sum_{i,j = 0,j > n}^{2n}\sum_{j,k = 0 }^n  p_{ijk}
\enq
Its information content becomes
\beq \label{inf}
\frac{\lp \frac{\partial \av{o}}{\partial \alpha} \rp^2 }{\av{o^{2}_\alpha(\delta)} - \av{o_\alpha(\delta)}^2} =  
\sum_{i = 0}^{2n} \sum_{j,k = 0}^n  p_{ijk}\:+ 2 \sum_{i,j = 0, j > n}^{2n}\sum_{k = 0}^n  p_{ijk}
\enq
where again we suppressed terms explicitly higher than $2n$. Since $p_{ijk} $ is symmetric in the last two indices, we can conclude that its information \eqref{inf} is simply
\beq
\sum_{i,j,k = 0}^{2n}p_{ijk} = \sum_{i+j+k = 2n} p_{ijk} + O(\sigma^{2n+1})
\enq
which is exactly the Fisher information content $\av{p  (\partial_\alpha \ln p)^2}$ of $p$ to order $2n$.
\section{Matrix approach to the information content and optimal observables} \label{pert. series}
We used in the main text the Edgeworth series of $\ln p(\delta)$ around the Gaussian distribution to derive our results. We describe in this appendix a method that is more involved though completely general, based uniquely on the explicit expressions of the moments. We discuss then why these two methods are in our case equivalent, irrespectively of the behavior of the function represented by the Edgeworth series.
\newline
\newline
This method is based on the expressions \eqref{poly} and \eqref{poly3},
which is the expansion of the information into the orthonormal system of polynomials, valid whenever the moment series determine uniquely $p(\delta)$. Expanding the polynomials in powers of $\delta$, we have
\beq
\begin{split}
P_n(\delta)& = \sum_{k = 0}^nC_{nk}\delta^k \\
s_n^\alpha & =  \sum_{k = 1}^nC_{nk}\frac{\partial m_k}{\partial \alpha}.
\end{split}
\enq
For each $N$, the triangular matrix $C_{nk}, n,k = 0,\cdots,N$ is the Cholesky factor of the inverse moment matrix of the same size,
\beq
C^TC  = M_N^{-1} , \quad \lb M_N \rb_{nk} = m_{n+k}, \quad n,k = 0,\cdots,N
\enq
Thus, if we obtain a perturbation series for the matrix $C$, we can find both $P_n(\delta)$ and $s_n^\alpha$. The optimal observable and Fisher information are then obtained by the relations \eqref{poly} and \eqref{poly3}. To obtain a perturbation series of the matrix $C$, we can proceed as follows. In a first step, consider that the total moment matrix is as in our case the sum $M = \bar M + \delta M$ of  a reference moment matrix $\bar M$, whose inverse has Cholesky factorisation $\bar M^{-1}= \bar C^T\bar C$, plus a perturbation $\delta M$. We can write
\beq
M^{-1} = \lp \bar M + \delta M \rp^{-1} =  \bar C^T \lp 1 +\bar C \delta M \bar C^T \rp^{-1}\bar C. 
\enq
If the perturbation is small enough follows
\beq \label{inverse}
M^{-1} = \bar C^T \sum_{k = 0}^\infty (-1)^k  A^k \bar C,
\enq
where we defined $A = \bar C\delta M \bar C^T$. Note that $A$ also has the following more insightful interpretation. It holds
\beq \label{pdfp}
A_{nm} = \av{\bar P_n \bar P_m} - \delta_{nm},
\enq
where the average is with respect to the true probability density function. This is easily shown using the fact that the reference polynomials are given by $\bar P_n(x) = \sum_k \bar C_{nk}\delta^k$. Thus, the matrix $A$ directly measures the deviations from orthogonality of the reference polynomials.
\newline
\newline
In a second step, we seek to find the matrix  $C$ associated to \eqref{inverse}. It is convenient to introduce the following notation, borrowed from \cite{Stewart:1997uq} who discusses the first order perturbation to the Cholesky decomposition. Given a matrix $A$, the matrix $L_{1/2}A$ is the lower triangular matrix obtained by zeroing the part above the diagonal, and multiplying by $1/2$ the diagonal itself. Likewise, $U_{1/2}A$ is the upper triangular matrix obtained by zeroing the entries below the diagonal and multiplying by $1/2$ the diagonal. If $A$ is symmetric we have that $\lb L_{1/2}A\rb^T  = U_{1/2}A$.  Putting formally
\beq \label{series}
C = \lp 1 + \sum_{k = 1}^\infty L_{1/2}Q^{(k)} \rp\bar C
\enq
for some yet unknown symmetric matrices $Q^{(k)}$, the requirement $C^TC = M^{-1}$ gives us from equation \eqref{inverse} the following relation at order $k$,
\beq
(-1)^kA^k = U_{1/2}Q^{(k)} + L_{1/2}Q^{(k)} + \sum_{i + j = k}U_{1/2}Q^{(i)}L_{1/2}Q^{(j)}
\enq
By definition of $L_{1/2}$ and $U_{1/2}$, we have $L_{1/2}Q^{(k)}  + U_{1/2}Q^{(k)} = Q^{(k)}$. We obtain thus the recursion relations
\beq
\begin{split}
Q^{(1)} &= -A \\
Q^{(k)} & = (-1)^kA^k - \sum_{i + j = k} U_{1/2} Q^{(i)}  L_{1/2} Q^{(j)}.
\end{split}
\enq
These relations together with \eqref{series} allow the perturbative calculation of the polynomials. We have the formal series
\beq
P_n(\delta) = \bar P_n(\delta) + \sum_{i = 1}^\infty \sum_{m = 0}^n L_{1/2} Q^{(i)}_{nm}\bar P_m(\delta).
\enq
Note that in general not all entries of the perturbation $\delta M$ are of the same order, so that the perturbation series must be further reorganized in order to obtain a consistent expansion.
\newline
\newline 
A connection to the Edgeworth series is the following. In our case, the reference distribution is the Gaussian with zero mean and variance $\sigma^2$. The orthonormal polynomials are the (suitably rescaled) Hermite polynomials
\beq \label{Po}
\bar P_n(\delta) = \frac{1}{\sqrt{n!}} H_n\lp\nu \rp.
\enq
It is then an interesting if somewhat lengthy exercise of algebra to derive the matrix $1+ A = \bar CM\bar C^T$, using only the explicit expression of the moments
\beq
m_n = \sum_{2k_2 + 3k_3 +\cdots = n} \sigma^{2n-2|\veck|} \prod_{i\ge 2} \lp \frac{S_i^{k_i}}{i!^{k_i} k_i!}\rp,
\enq
(with the understanding that $S_2 = 1$, and $|\veck| = \sum_i k_i$) and that of the Hermite polynomials
\beq
H_n(\nu) = n!\sum_{2k \le n}\frac{(-1)^k}{k!(n-2k)!}\frac{\nu^{n-2k}}{2^k}.
\enq
The result can be written as follows
\beq\label{Amatrix}
\lp \bar CM\bar C^T \rp_{ij} = \frac{1}{\sqrt{i!j!}}\sum_{n \ge 0} \sigma^n \sum_{k_3,k_4,\cdots} (i;\:j;\:n + 2|\veck|) \prod_{p \ge 3}\lp \frac{S_p^{k_p}}{p!^{k_p} k_p !}\rp,
\enq
where the sum runs over all set positive integers $(k_3,k_4 \cdots)$ with the condition $\sum (i-2)k_i = n$, and the symbol $(i;\:j;\:k)$ is the integral of three Hermite polynomials with respect to the Gaussian distribution of unit variance,
\beq \label{reprs}
(i;\:j;\:k) = \av{H_i(\nu)H_j(\nu)H_k(\nu)}_G. 
\enq
In this representation, it can be readily verified that all sums contain only a finite number of terms and thus do not suffer any ambiguities. This is because this integral of three Hermite polynomials satisfy the triangle conditions, non zero only for $|i-j| \le k \le |i + j|$ at fixed $i$ and $j$. The values of $n$ covers only a finite range at each $i$ and $j$. However, using the integral representation \eqref{reprs} of the symbols, one can try put the full sums under the integral $\av{\cdots}_G$, getting precisely
\beq \label{Amtrix}
\lp \bar CM\bar C^T \rp_{ij} = \frac{1}{\sqrt{i!j!}}\av{H_i(\nu) H_j(\nu) \lp 1 + \Delta p(\sigma,\nu)\rp}_G
\enq
where $\Delta  p$ is the Edgeworth series as in equation \eqref{Edgeworth}. According to \eqref{pdfp}, $\lb \bar CM\bar C^T \rb_{ij}$ must be equal to $\av {\bar H_i\bar H_j}/\sqrt{i!j!}$, where the average is with respect to the true $p(\delta)$, or a $p(\delta)$ with the correct series of moments when this is not unique. Comparing with $\eqref{Amtrix}$, we have thus rederived here the Edgeworth representation of $p(\delta)$ in the case of convergence of that series. If the series for $\Delta p$ does not converge, then the representation \eqref{Amtrix} does not make sense, but the exact \eqref{Amatrix} still does, coinciding with the formal expansion of the Edgeworth series in \eqref{Amtrix}. This justifies the use of the Edgeworth series in this work, since it reproduces the correct perturbation series in all cases.
\section{Edgeworth series and information} \label{appendixF}
We provided in the text the first two terms of the Edgeworth series of $\ln p$, allowing the calculation of the Fisher information including $\sigma^2$, and of the observable capturing the information including $\sigma^4$. We list here for completeness the next two terms of $\ln p$ as well, allowing us to read out the next two terms in the optimal observables and the next term in the total information. We then obtain the Taylor expansion of the information dominant observable $f_0(\delta)$ at all orders. We work for simplicity within the hierarchical model. Loop corrections would enter $g_5(\nu)$ and $g_6(\nu)$. The corresponding adaptations of $f_4(\delta)$ and of the $\sigma^4$ terms of $F$ would be required if their inclusion is desired.
\newline
\newline
Viewed as a power series in $\sigma$, the relation between $1 + \Delta p$ in equation \eqref{Edgeworth} and $\ln p$ is the same as the relation between a moment generating function and a cumulant generating function. The first four terms of $\Delta p$ reads
\beq \nonumber
\begin{split}
\Delta p &= \sigma \frac{S_3}{6} H_3 + \sigma^2 \lp\frac{S_4}{24} + \frac{S_3^2}{72}H_6 \rp \\
&  +\sigma^3 \lp \frac{S_3^4}{1296}H_9 + \frac{S_3S_4}{144}H_7 + \frac{S_5}{120}H_5\rp \\
&+ \sigma^4 \lp \frac{S_3^4}{31104}H_{12} + \frac{S_3^2S_4}{1728}H_{10} + \frac{S_4^2}{1152}H_8 + \frac{S_5S_3}{720}H_8 + \frac{S_6}{720}H_6 \rp.
\end{split}
\enq
From this, we get
\beq \nonumber
\ln p = -\frac{\nu^2}{2} - \frac 12 \ln \lp 2\pi \sigma^2\rp + \sum_{n = 1}^4\sigma^n g_{2 + n}(\nu) + O(\sigma^5),
\enq
with
\beq
\begin{split} \label{gp}
g_3(\nu) 
= \frac{S_3}{6}&\lp \nu^3 -1\rp \\
g_4(\nu) 
 =  \frac{S_4}{24}& \lp \nu^4 - 6\nu^2 + 3 \rp - \frac{S_3^2}{24} \lp 3\nu^4 - 12\nu^2 +5 \rp\\
g_5(\nu)
 = \frac{S_5}{120}&\lp \nu^5 - 10\nu^3 + 15\nu \rp - \frac{S_3S_4}{12} \lp \nu^5 -7\nu^3 + 8\nu\rp \\
+ \frac{S_3^3}{24} &\lp3\nu^5 -16\nu^3 + 15\nu \rp \\
g_6(\nu) 
 = \frac{S_6}{720}&\lp \nu^6 - 15\nu^4 + 45\nu^2 -15\rp \\
 \:- \frac{S_5S_3}{48} &\lp \nu^6 -11\nu^4  + 25\nu^2 -7\rp \\
 \: - \frac{S_4^2}{144}&\lp 2\nu^6 -21 \nu^4 + 48\nu^2 -12 \rp \\
 \:+ \frac{S_3^2S_4}{48}& \lp 7\nu^6 -59 \nu^4 + 109\nu^2 -25 \rp \\
 \: - \frac{S_3^4}{48} &\lp 7\nu^6 -48\nu^4 + 75\nu^2 -15\rp,
\end{split}
\enq
in perfect agreement with \cite{Takemura}.  From these polynomials we read out the optimal observable
\beq
o(\delta) = f_0(\delta) + \sigma^4f_4(\delta)
\enq
where $f_0 = \delta^2 + \sum_{n = 3}^6a_n \delta^n$ is given explicitly in the main text, and
\beq
\begin{split}
f_4(\delta)& = \delta \lp  \frac{S_5}{4} - \frac 43 S_3S_4 + \frac 54 S_3^2\rp  \\
&\:+ \delta^2\lp \frac{S_6}{8} - \frac{25}{24}S_5S_3 - \frac 23 S_4^2 + \frac{109}{24}S_4S_3^2 -\frac{25}{8}S_3^3\rp
\end{split}
\enq
capturing the entire information including $\sigma^8$ and lower order terms. On the other hand, we can obtain the total information from these polynomials including $\sigma^4$. Separating the independent contribution from each moment
\beq
F = F_2 + F_3 + F_4 + O(\sigma^6),\quad F_n = s_n^2,
\enq
we get
\beq
\begin{split}
F_2 &= \frac 12  -  \frac{\sigma^2}4 \lp S_4-S_3^3 \rp + \frac{\sigma^4}{8} \lp S_4-S_3^2 \rp^2  +O(\sigma^6)\\
F_3 &=  \sigma^2 \frac{S_3^2}{6}  + \sigma^4 \lp  \frac{S_3S_5}{6} -\frac{11}{12}S_3^2S_4 + \frac 34 S_3^4 \rp +O(\sigma^6) \\
F_4 &= \frac{\sigma^4}{24} \lp S_4 - 3S_3^2\rp^2 + O(\sigma^6)
\end{split}
\enq
We can give explicitly the leading term of $F_n$, simply from reading the leading coefficients of the $g$ polynomials,
\beq \label{claim}
F_n = n! \:G^2_{nn }\:\sigma^{2n-4} + O(\sigma^{2n-2)}.
\enq
This follows from comparing $\eqref{lnp}$ to its exact form $\sum_n s_n P_n(\nu)$, as in point (ii) in section \ref{sections}. The leading term of $P_n(\nu)$ is $H_n(\nu) /\sqrt{n!}$ (see \eqref{Po}). Since $H_n(\nu) = \nu^n +\cdots$ we infer
\beq
s_n = \sqrt{n!}R_{nn}\sigma^{n-2} + O(\sigma^{n}).
\enq
From $R_{nn} = -G_{nn}$ and $F_n = s_n^2$ follows our claim \eqref{claim}.
We now turn to the derivation of  the variance independent, loop corrections independent $f_0(\delta) = \sum_{n \ge 2}a_n\delta^n$. From our results in the main text, we have
\beq
a_n = -2G_{nn},
\enq
where $G_{nn}$ is the leading coefficient of the polynomial $g_n(\nu)$ entering the Edgeworth series of $\ln p$. Again, we can make use of the results of \cite{Takemura}, who obtained in their equations (2.25) and (2.34) the necessary ingredients. In their notation and conventions, their $\beta_i$ corresponds for us to
\beq \label{beta}
 \beta_i = \sigma^{i-2}\frac{S_i}{i!}.
\enq
They derive that the leading coefficient of the polynomials accompanying $\beta_{3}^{k_3}\beta_{4}^{k_4} \cdots$ is given by (writing $\veck = (k_3,k_4,\cdots))$
\beq
c(\veck) = (-1)^{|\veck| -1}\lp \prod_{i \ge 3} i^{k_i} \rp \prod_{j= 0}^{|\veck| -3} \lp \sum_{i\ge 3} (i-1)k_i  - j\rp,
\enq
where the product is unity if $|\veck|  = 2$. If $|\veck |= 1$ the leading coefficient is 1. The power of $\sigma$ accompanying this term is according to \eqref{beta} equal to $\sum_{i \ge 3}(i-2)k_i $. In the series for $\ln p$, the polynomial $g_n$ multiply $\sigma^{n-2}$. Fixing thus the order $n-2$, we get that the leading coefficient of $g_n$ is given by
\beq
G_{nn} = \sum_{\veck} \prod_{i \ge 3}\lp \frac{S_i^{k_i}}{i!^{k_i}k_i!} \rp c(\veck),
\enq
where the sum includes all vectors of positive integers $\veck$ such that
\beq
\sum_{i\ge3} (i-2)k_i = n-2. 
\enq
Elementary manipulations leads to
\beq
\begin{split}
G_{nn} = \frac 1 {n!}\sum_{\veck} (-)^{|\veck| -1}\lp n-2 + |\veck|\rp! 
\prod_{i \ge 3}\lp \frac{S_i^{k_i}}{\lp i-1\rp!^{k_i}k_i!} \rp,
\end{split}
\enq
Equation \eqref{f0} follows.
\end{appendix}

\end{document}